\documentclass[9pt,onehalfspacing]{elife}

\title{COVID-19 infectivity profile correction}

\author[1*]{Peter Ashcroft}
\author[1]{Jana S. Huisman}
\author[1]{Sonja Lehtinen}
\author[1]{Judith A. Bouman}
\author[2]{Christian L. Althaus}
\author[1]{Roland R. Regoes}
\author[1*]{Sebastian Bonhoeffer}
\affil[1]{Institute of Integrative Biology, ETH Z\"urich, Z\"urich, Switzerland}
\affil[2]{Institute of Social and Preventive Medicine, University of Bern, Bern, Switzerland}

\corr{peter.ashcroft@env.ethz.ch}{PA}
\corr{seb@env.ethz.ch}{SB}

\begin{document}

\maketitle

\begin{abstract}
The infectivity profile of an individual with COVID-19 is attributed to the paper \emph{Temporal dynamics in viral shedding and transmissibility of COVID-19} by He \emph{et al.}, published in Nature Medicine in April 2020.
However, the analysis within this paper contains a mistake such that the published infectivity profile is incorrect and the conclusion that infectiousness begins 2.3 days before symptom onset is no longer supported.
In this document we discuss the error and compute the correct infectivity profile.
We also establish confidence intervals on this profile, quantify the difference between the published and the corrected profiles, and discuss an issue of normalisation when fitting serial interval data.
This infectivity profile plays a central role in policy and decision making, thus it is crucial that this issue is corrected with the utmost urgency to prevent the propagation of this error into further studies and policies.
We hope that this preprint will reach all researchers and policy makers who are using the incorrect infectivity profile to inform their work.
\end{abstract}

\section{Introduction}
While investigating the results of the paper \emph{Temporal dynamics in viral shedding and transmissibility of COVID-19} \citep{he:NatMed:2020}, we have found an erroneous step in the likelihood calculation which is cause for concern.
The consequence of this step is that two datapoints with negative serial interval are dropped from the calculation, without any explicit mentioning in the text of the manuscript.
The inclusion of these datapoints results in an infectiousness profile that is substantially different from the one shown in Figure 1C of your original publication.
As a result, the infectiousness starts significantly before the reported $2.3$ days before the onset of symptoms.
We still find, however, a presymptomatic infection fraction of $\sim 45\%$ in agreement with the conclusion of \cite{he:NatMed:2020}.
Given that the estimate of $2.3$ days of infectiousness before symptom onset is highly relevant to the implementation of contact tracing, we believe it is of very high importance to clarify this situation.
Our reanalysis suggests that tracing contacts of infected index cases as far back as 2 or 3 days before symptom onset in the index case might not be sufficient to find all secondary cases.
In addition, we remark on a less consequential issue with the normalisation of the likelihood, which awards higher weight to transmission pairs with more uncertain symptom onset times of the index case, but does not affect the results significantly.
Due to the central position this study currently has in the field, there is a high probability that these errors propagate in future studies.
Therefore, a fast response towards this issue is crucial.
We note that detecting this issue was only possible thanks to the availability and accessibility of the code and data that accompany the publication.

With this letter we address the following three points:
\begin{enumerate}
\item The infectivity profile is computed without erroneously dropping datapoints and we compare this corrected profile with the published profile;
\item Confidence intervals via likelihood profiling are provided for the infectivity profile;
\item An issue relating to the normalisation of the likelihood over serial interval ranges is discussed.
\end{enumerate}

\section{Results}

\begin{figure}[h]
\centering
\includegraphics[width = \textwidth]{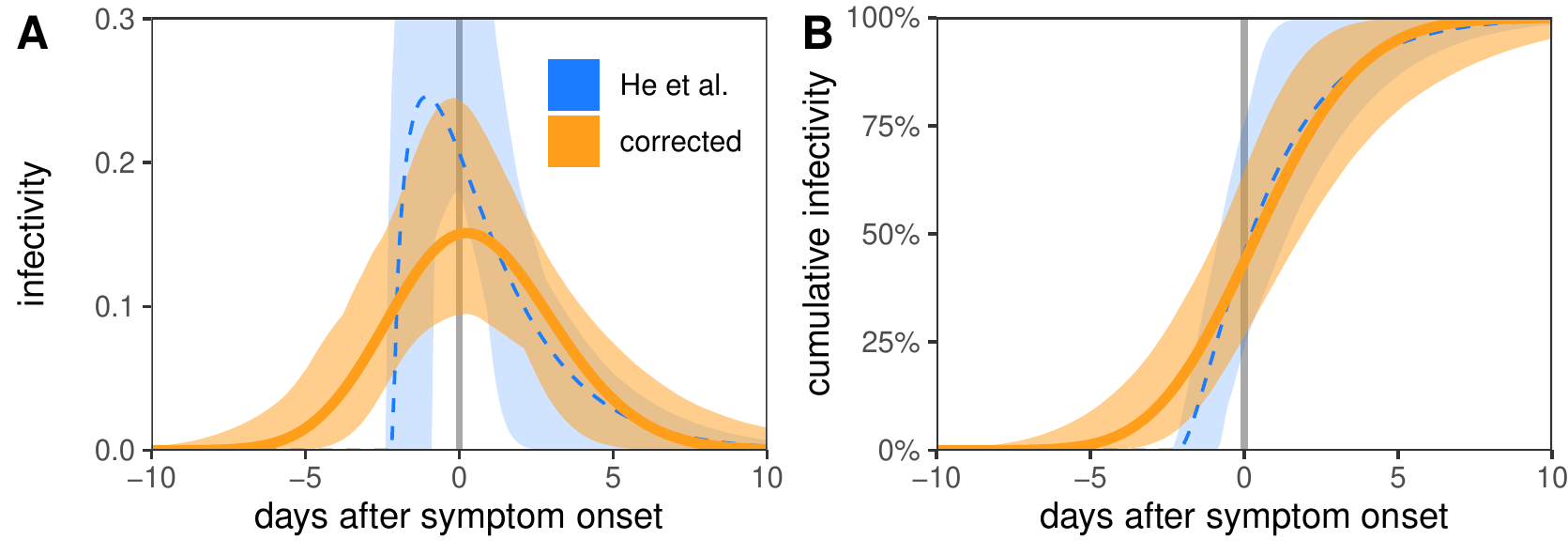}
\caption{
The infectivity profiles extracted from the serial interval data and the log-normally distributed incubation period, as performed in \cite{he:NatMed:2020}. Here we use an adaptive grid search to reconstruct the likelihood landscape over the three parameters of the shifted gamma distribution.
The confidence intervals are the range of the infectivity profiles which have a likelihood-ratio test statistic within $95\%$ of a $\chi^2$ distribution with 3 degrees of freedom when compared to the maximum likelihood estimate.
Panel A shows the probability density function and is analogous to Figure 1C of \cite{he:NatMed:2020}.
The blue dashed line is the maximum likelihood estimate using the method from \cite{he:NatMed:2020},
while the solid orange line is the corrected maximum likelihood estimate.
Panel B shows the corresponding cumulative density functions.
}
\label{fig:gamma}
\end{figure}

The infectivity profile, $p(t)$, describes the infectiousness of an individual at a time $t$ relative to the onset of their symptoms.
When this is convolved with an incubation period distribution $g(t)$ [from \cite{li:NEJM:2020}], one recovers the serial interval distribution, describing the time between symptom onsets in a transmission pair.
This approach was used in \citep{he:NatMed:2020}, with a fixed incubation period distribution and empirical serial interval distribution, to infer the infectivity profile for COVID-19.

The optimisation procedure maximises the likelihood of observing the empirical serial interval distribution under a model, which is specified by the parameters of the infectivity profile.
This profile is parametrised as a shifted gamma distribution.
Full details of this procedure can be found in \cite{he:NatMed:2020}. 

In the fitting procedure used in the script \texttt{Fig1c\_Rscript.R} (available at \url{https://github.com/ehylau/COVID-19}), the following condition is used in the return line of the likelihood function:
\begin{verbatim}
return(-sum(lli[!is.infinite(lli)]))
\end{verbatim}
This condition will erroneously drop any datapoint that has a probability of zero (and hence a log-probability of $-\infty$) under the current model parameters.
As the optimisation is initiated with a shift value of $2.5$ days, two datapoints (54 and 68) are dropped from the beginning of the fit procedure.
This then leads to an erroneous maximum likelihood infectiousness profile, which is displayed in Figure~1C of the original manuscript \citep{he:NatMed:2020}.
Initiating the fitting procedure at a shift of $4$ days shows convergence to a very different optimum infectiousness profile.

Here we use an adaptive grid search algorithm to scan the three-dimensional parameter space of the shifted gamma distribution which describes the infectiousness profile.
We compute the log-likelihood with and without the return condition in the likelihood function at each point in parameter space to construct likelihood surfaces.
The maximum likelihood parameter values that we find are enumerated in Table~\ref{tab:mle}.

\begin{table}[h]
\begin{tabular}{c r r}
\toprule
parameter & He \emph{et al.} & corrected \\
\midrule
shape & 1.56250 & 97.18750 \\
rate & 0.53125 & 3.71875 \\
shift (days) & 2.12500 & 25.62500 \\
\bottomrule
\end{tabular}
\caption{Maximum likelihood parameter estimates based on our adaptive grid search approach using the method of \citep{he:NatMed:2020} and the corrected computation.}
\label{tab:mle}
\end{table}

We construct confidence intervals around the distribution via likelihood ratio tests, compared to the maximum likelihood estimate (also known as likelihood profiling).
This leads to the optimum infectivity profiles and confidence intervals shown in Figure~\ref{fig:gamma}.
We also find a presymptomatic infection fraction of 45.6\% [23.8\%,75.8\%] using the \cite{he:NatMed:2020} method and 43.7\% [26.4\%,64.5\%] using the corrected profile, where numbers in square brackets represent the 95\% confidence interval.

\begin{figure}
\centering
\includegraphics[width = 0.6\textwidth]{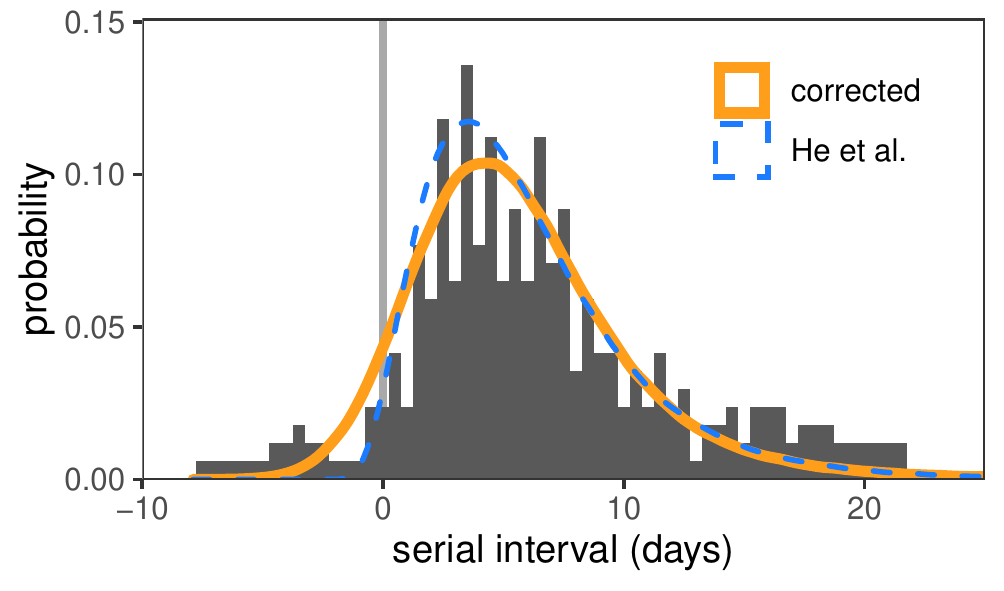}
\caption{Using the maximum likelihood estimates of the infectivity profile from Fig. \ref{fig:serial-fit} we can reconstruct the serial interval. 
We sample $10^6$ infection times from the infectivity profile, and add this to samples from the log-normally distributed incubation period to generate samples of the serial interval.
We then plot the probability density of these serial intervals (filled density profiles).
We compare this to the serial interval data used in \cite{he:NatMed:2020}, where we have added points for each day in the possible serial interval range.
}
\label{fig:serial-fit}
\end{figure}

The correct optimum fits in Figure~\ref{fig:gamma} are smoother than the ones which drop the data points.
Although there is some asymmetry in the profiles within the confidence interval, the correct optimum solution has a very large shape parameter and approaches a normal distribution.
We can also use these fitted distributions to reconstruct the serial interval (Fig. \ref{fig:serial-fit}), the distribution of which is broader when all datapoints are taken into account.

To further quantify the difference between the published and corrected infectivity profiles, we can use an example based on contact tracing.
We use the infectivity profiles to answer the following question:
What fraction of presymptomatic infections are traced if we look back $t$ days from symptom onset?
Formally, this fraction is defined as
\begin{equation}
f(t) = \dfrac{\int_{-t}^0 p(t'){\rm d}t'}{\int_{-\infty}^0 p(t'){\rm d}t'}.
\label{eq:f}
\end{equation}
These values are enumerated in Table~\ref{tab:f}.
We see that while the published infectivity profile suggests $98\%$ of presymptomatic infections occur within two days symptom onset, the corrected distribution suggests on $61\%$ of presymptomatic infections will be traced.
Thus the published profile overestimates the efficacy of contact tracing, while the corrected distribution tells us we need to look back at least 4 days to catch 90\% of presymptomatic infections.

\begin{table}[h]
\begin{tabular}{r r r}
\toprule
time (days) & He \emph{et al.} & corrected \\
\midrule
1 & 50\% [37\%,100\%] & 33\% [19\%,51\%] \\
2 & 98\% [87\%,100\%] & 61\% [40\%,83\%] \\
3 & 100\% [100\%,100\%] & 80\% [57\%,96\%] \\
4 & 100\% [100\%,100\%] & 91\% [71\%,99\%] \\
5 & 100\% [100\%,100\%] & 97\% [82\%,100\%] \\
\bottomrule
\end{tabular}
\caption{The fraction of presymptomatic infections that are traced if we look back $t$ days from symptom onset using the published and corrected infectivity profiles. The computed quantity $f(t)$ is described in Eq.~\ref{eq:f}. Values in brackets represent 95\% confidence intervals of $f(t)$ when accounting for the uncertainty in the infectivity profiles.}
\label{tab:f}
\end{table}

A second less-consequential problem in the methodology of \cite{he:NatMed:2020} is that a normalisation factor is missing in the likelihood function when considering transmission pairs with serial interval estimates specified by a range.
Ignoring this normalisation awards higher weight to transmission pairs with more uncertain symptom onset times of the index case.

Concretely, the probability under model $\theta$ to observe a window of symptom onset of the index case $(t_{S1l},t_{S1u})$ and symptom onset in the secondary case on day $t_{S2}$ is defined in the original manuscript as
\begin{equation}
L(t_{S1u},t_{S1l},t_{S2} | \theta) = \int_{t_{S1l}}^{t_{S1u}} \int_{-\infty}^{t_{S2}} p(t_I - t_{S1}) g(t_{S2} - t_I) {\rm d}t_I\, {\rm d}t_{S1},
\end{equation}
where $p(t)$ is the infectivity profile and $g(t)$ is the incubation period distribution.
The outer integral over the symptom onset window of the index case should include an accompanying probability to observe the onset time $t_{S1}$, ${\rm Pr}(t_{S1})$.
I.e.
\begin{equation}
L(t_{S1u},t_{S1l},t_{S2} | \theta) = \dfrac{\int_{t_{S1l}}^{t_{S1u}} {\rm Pr}(t_{S1}) \int_{-\infty}^{t_{S2}} p(t_I - t_{S1}) g(t_{S2} - t_I) {\rm d}t_I\, {\rm d}t_{S1}}{\int_{t_{S1l}}^{t_{S1u}} {\rm Pr}(t_{S1}){\rm d}t_{S1}}.
\end{equation}
Assuming a uniform distribution for ${\rm Pr}(t_{S1})$, this simplifies to
\begin{equation}
L(t_{S1u},t_{S1l},t_{S2} | \theta) = \frac{1}{t_{S1u}-t_{S1l}}\int_{t_{S1l}}^{t_{S1u}} \int_{-\infty}^{t_{S2}} p(t_I - t_{S1}) g(t_{S2} - t_I) {\rm d}t_I\, {\rm d}t_{S1}.
\end{equation}
We find that including this normalisation has little effect on the location of the optimum fit.
However, these full details should have been included in the optimisation procedure.

\section{Footnotes}
The code used to generate these results is archived at \url{https://zenodo.org/badge/latestdoi/278170144}.



%
%
%
%
%
%
\end{document}